\documentclass[superscriptaddress,reprint,aps,prl,amsfonts,amsmath]{revtex4-1}
\usepackage{amsmath, amssymb, amsfonts}
\usepackage{graphicx,epstopdf,bm}
\usepackage[colorlinks,bookmarks,urlcolor=blue,citecolor=blue,linkcolor=blue]{hyperref}
\usepackage[T1]{fontenc}
\usepackage{color}

\usepackage{soul}
\usepackage{bbm} %$\mathbbm{abcdefghijklmnopqrstuvwxyz}$
 \newcommand{\pd}[2]{\partial_{#2} #1 }
\newcommand{\pdd}[2]{\partial^{2}_{#2} #1 }

\newcommand{\abs}[1]{\left\vert #1 \right\vert}

\newcommand{\explr}[1]{\exp\left[ #1 \right]}

\newcommand{\qedsymbol}{\vbox{\hrule height0.6pt\hbox{\vrule height1.3ex width0.6pt\hskip0.8ex\vrule width0.6pt}\hrule height0.6pt}}

\newcommand{\en}{n}
\newcommand{\tn}{n_{r}}

\newcommand{\dif}{\frac{\sigma^{2}}{2}}

\begin{document}
\title{Effects of moderate noise on a limit cycle oscillator: Counter rotation and bistability}
\author{Jay M. Newby}
\affiliation{Mathematical Bioscience Institute, Ohio State University, 1735 Neil Ave. Columbus, OH 43210}
\author{Michael A. Schwemmer}
\affiliation{Mathematical Bioscience Institute, Ohio State University, 1735 Neil Ave. Columbus, OH 43210}

\begin{abstract}
 The effects of noise on the dynamics of nonlinear systems is known to lead to many counter-intuitive behaviors. 
Using simple planar limit cycle oscillators, we show that the addition of moderate noise leads to qualitatively different dynamics. 
In particular, the system can appear bistable, rotate in the opposite direction of the deterministic limit cycle, or cease oscillating altogether. 
Utilizing standard techniques from stochastic calculus and recently developed stochastic phase reduction methods, we elucidate the mechanisms underlying the different dynamics and verify our analysis with the use of numerical simulations. 
Lastly, we show that similar bistable behavior is found when moderate noise is applied to the more biologically realistic FitzHugh-Nagumo model. 
 \end{abstract}

%;% isochrones => ito phase drift
%;% ito phase drift => fick phase drift (through SDE <=> FP equation)
%;% fick drift => stationary density
%;% ito drift => average jump at a given point in time
%;% fick drift => fraction of time spent in the neighborhood of a point relative to other points as t -> \infty
%;% 
%;% radial symmetry => fick drift = ito drift. Hence, isochrones => stationary density directly.
%;% no radial symmetry => fick drift != ito drift. Hence, bistability cannot be DIRECTLY inferred from isochrones.

\maketitle
 Limit cycle oscillators have been widely used to model various natural phenomena \cite{kuramoto,syncbook,winfree}.
As such, they have been the subject of extensive study in the field of nonlinear science.
In recent years, the effects of noise on the dynamics of limit cycle oscillators has received much interest \cite{bolandetal2009,gonzeetal2002,koeppletal2011,teramae2009,yoshimura2008}. 
When the noise is weak, formal reduction methods can be performed to reduce the dimensionality of the system to a so-called phase equation \cite{teramae2009, yoshimura2008}.  
In this case, one can analytically show that both the magnitude (and correlation time for colored noise) of the added noise shift the mean frequency of oscillations away from the natural frequency of the limit cycle.
On the other hand, when the noise is large, the trajectories of the system appear completely random and bear no resemblance to the deterministic limit cycle behavior.

The case of moderate noise applied to a limit cycle oscillator has received less attention. 
It is known that moderate noise can cause stochastic resonance in systems close to a bifurcation to limit cycle oscillations \cite{gangetal1993,rappelandstrogatz1994}, and in systems where limit cycles arise as the result of periodic forcing \cite{bolandetal2009}.  However, an exploration of how moderate noise interacts with the underlying deterministic dynamics of a system displaying limit cycle behavior has not yet been undertaken and is the purpose of the current Letter.
We find that an oscillator subject to moderate noise can display numerous interesting and counter-intuitive behaviors.
In some cases, the addition of noise causes the phase to behave like a bistable switch, while in other cases noise can act to completely eliminate oscillations, and even cause the trajectories to rotate in the {\it opposite} direction of the deterministic limit cycle.  
% We illustrate these effects using simple planar models displaying limit cycle behavior.  
% The occurrence of bistability and the elimination of oscillations in limit cycle systems has been previously reported.

Bistability in the amplitude of the limit cycle has been shown to occur in the Stuart--Landau (SL) system when it is subjected to periodic forcing \cite{legaletal2001}, or specially constructed stochastic forcing \cite{staliunasetal2009}.
It is also known that coupling limit cycle oscillators together can eliminate oscillations \cite{ermentroutandkopell1990,saxenaetal2012}.  % through a phenomenon known as amplitude death
However, we show that these phenomena can occur in a planar limit cycle system when each component is subjected to additive white noise.
% Using standard techniques from stochastic calculus \cite{gardiner} and recently developed stochastic phase reduction methods \cite{teramae2009, yoshimura2008}, we also illustrate that the mechanisms underlying the phenomena in this case are completely different than those previously discovered.
% Furthermore, to our knowledge, we are not aware of any situation where the trajectories of an oscillatory system were caused to rotate in the opposite direction of the limit cycle.
% This will certainly lead to interesting synchronization phenomena when the oscillators are coupled, e.g. \cite{bhowmicketal2011}.

For simplicity, we consider a radially symmetric deterministic system. 
One such symmetric oscillator that we employ is the SL oscillator \cite{stuart1960}, which has been used to model the shedding of vortices in the two-dimensional wake of a cylinder at low Reynolds number (e.g., \cite{legaletal2001,provansal1987}). 
We first add noise so that the radial symmetry is statistically preserved, and then consider the system when the symmetry is broken.
 % and show that the oscillator behaves like a bistable stochastic process.
Finally, we show that similar behavior is seen in the non-radially symmetric, FitzHugh--Nagumo (FHN) oscillator \cite{fitzhugh1961,nagumo1962} with additive white noise.

\begin{figure}[b]
  \centering
  \includegraphics[width=8cm]{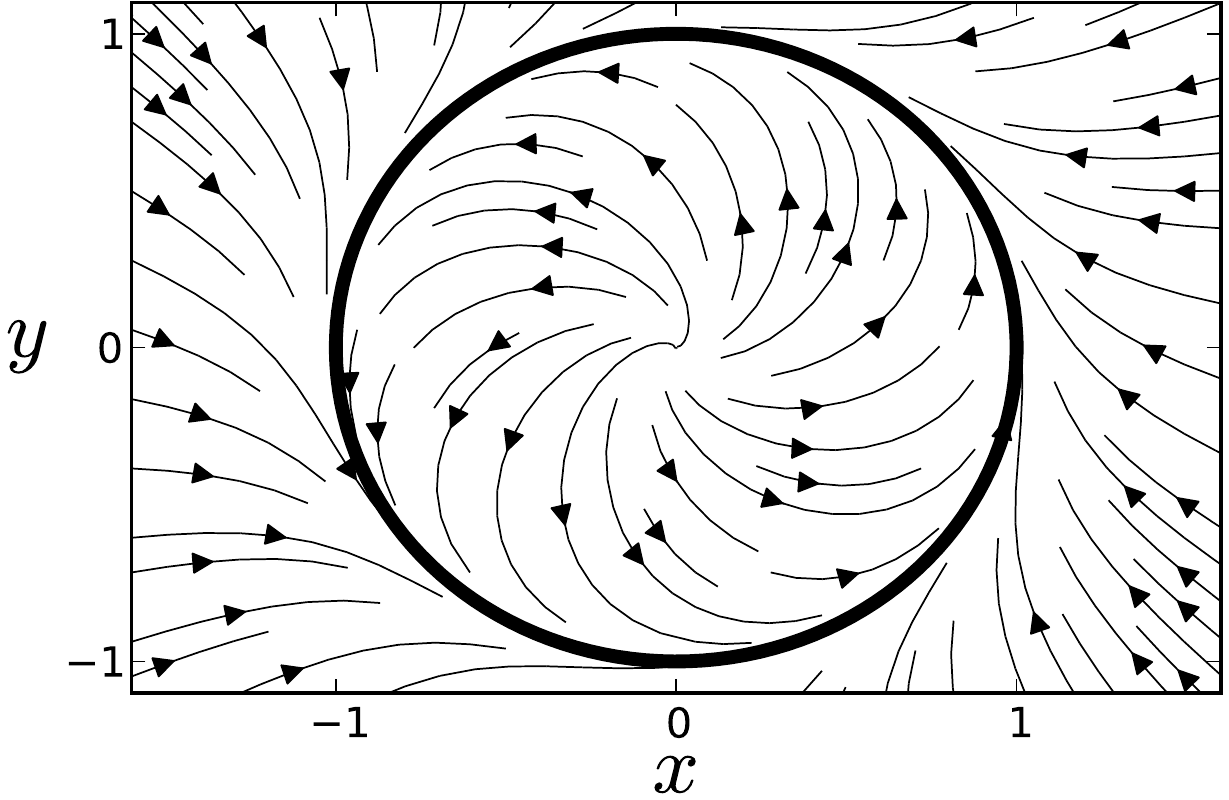}
  \caption{Radial isochrone clock model (equation (\ref{eq:5}) with $c = 0$), with stream lines indicating the direction of the vector field.}
  \label{fig:intro}
\end{figure}

Consider the following deterministic oscillator in polar coordinates
\begin{equation}
  \label{eq:5}
  \dot{\theta} = \omega - \gamma c Q(\rho), \quad   \dot{\rho} = -\gamma \rho(\rho^{2} - 1) % V1 Q = \rho^{2} - 1, V2 Q = -(1-\rho)^{2},
\end{equation} 
where the function $Q(\rho)$ is such that $Q(1) = 0$ and determines the rotation away from the limit cycle.
We assume that the limit cycle is strongly attracting so that $\gamma \gg \omega$ is a large parameter.
If $c = 0$, then \eqref{eq:5} is independent of $Q$ and yields the radial isochrone clock model when $\gamma=1$ (see Fig.~\ref{fig:intro}).  In this case, the rotation $\dot{\theta}$ is constant away from the limit cycle and independent of $\rho$.
If $c \neq 0$, then the rotation changes direction for values of $\rho=\rho_{*}$ such that $Q(\rho_{*}) = \omega/(c\gamma)$.
We consider the following two cases:
\begin{equation}
  \label{eq:6}
  Q_{1}(\rho) = \rho^{2} - 1, \quad Q_{2}(\rho) = - \omega(1 - \rho)^{2}.
\end{equation}
(Note that $Q_{1}$ gives the SL system.)
It follows that the limit cycle rotates counter clockwise on $\rho = 1$.
\begin{figure*}[tb]
  \centering
  \includegraphics[width=18cm]{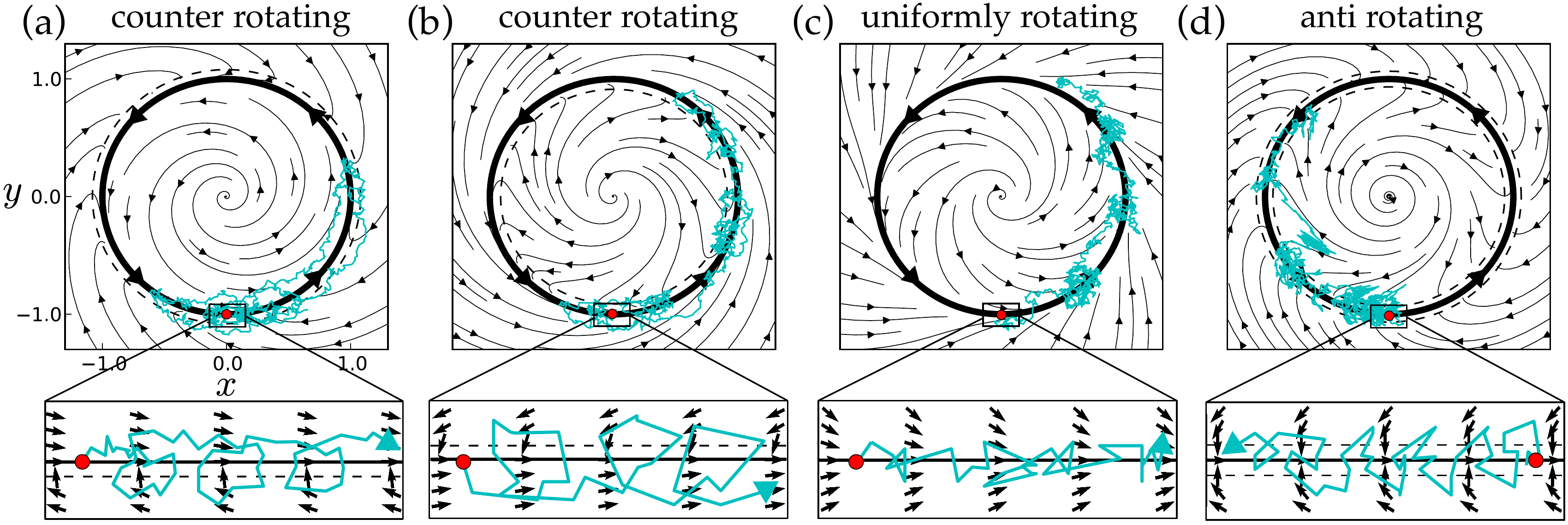}
  \caption{Deterministic phase plane for $c\neq 0$.  Light blue curves are short (approximately half of one period) numerically simulated stochastic trajectories of (\ref{eq:14}) starting at $(x,y) = (0, -1)$ (red dot). Thin black curves are streamlines of the deterministic vector field. (a, b) counter rotating (using $Q_{1}$) with (a) $c = 4$ and (b) $c = -4$. (c, d) using $Q_{2}$ with (c) uniformly rotating $c=4$ and (d) anti-rotating with $c = -15$. The lower panels are close-up sketches of a stochastic trajectory near the indicated part of the limit cycle.}
  \label{fig:cases}
\end{figure*}
Then, depending on the sign of $c$, there are four cases for rotation depicted in Fig.~\ref{fig:cases}. 
For $Q_{1}$ (Fig.~\ref{fig:cases}a,b) there is a unique value of $\rho_{*} = \sqrt{1+\omega/(c\gamma)}$ where the rotation changes sign; if $c > 0$ then $\rho_{*}>1$ and if $c < 0$ then $\rho_{*} <1$.
We refer to this case as {\em counter rotating}.
On the other hand, for $Q_{2}$ (Fig.~\ref{fig:cases}c,d) we have that $\rho_{*} = 1 \pm \sqrt{1/(-c\gamma)}$. %;% omega term
Hence, there are two values of $\rho_{*}$ if $c<0$ (one inside the limit cycle and one outside) and none if $c>0$.
We refer to this case as {\em uniformly rotating} for $c>0$ and {\em anti rotating} for $c<0$.
In both cases, $\rho_{*}\to 1$ as $\gamma \to \infty$.
% Hence, fluctuations that are large enough to move the trajectory across $\rho_{*}$,  on average, rotate in the opposite direction of the limit cycle.

Consider, the stochastic system with independent additive noise in cartesian coordinates given by
\begin{align}
  \label{eq:14}
  \dot{x} &= -\omega y + \gamma x(1-\rho^{2}) + c \gamma y Q(\rho) + \sigma G_x \xi_{x}(t) \\
  \dot{y} &= \omega x + \gamma y(1-\rho^{2}) - c\gamma x Q(\rho) + \sigma G_y \xi_{y}(t), \nonumber
\end{align}
where $\rho = \sqrt{x^{2}+y^{2}}$ and $G_{x}^{2} + G_{y}^{2} = 1$. 
(Note all stochastic simulations are performed using the Euler-Maruyama method).
If $G_{x} = G_{y}$, then the stochastic process is rotationally symmetric, and the stationary density can be computed exactly.

We transform the system from $(x, y)\to (\varphi, r)$, where $\varphi$ is the asymptotic phase \cite{kuramoto} and the amplitude $r$ is the distance from the limit cycle.
Lines of constant $\varphi$ are called {\em isochrones}; all deterministic trajectories starting on an isochrone converge as $t\to\infty$.
Hence, isochrones encode information about how the phase of a trajectory responds to a perturbation away from the limit cycle.
The transformation is given by
\begin{equation}
  r = \sqrt{x^2+y^2}-1,\quad \varphi  = \frac{1} {\omega} \left[\tan^{-1}(y/x) + c H(r) \right], % \ln(x^2+y^2)/2
\end{equation}
where $ H(r) \equiv \int_{0}^{r}\frac{Q(r'+1)dr'}{f(r')}$, $f(r) \equiv -r(r+1)(r+2)$.
% \begin{equation}
%   \label{eq:12}
%   H(r) \equiv \int_{0}^{r}\frac{Q(r'+1)dr'}{f(r')},\quad f(r) \equiv r(r+1)(r+2)
% \end{equation}
As such, the limit cycle occurs at $r=0$.
The inverse transformation is
\begin{equation}
  x = (r+1)\cos(\alpha(\varphi,r)),\quad  y = (r+1)\sin(\alpha(\varphi,r)),
\end{equation}
where $\alpha(\varphi,r) \equiv \omega\varphi + cH(r)$.
To transform \eqref{eq:14} into $(\varphi, r)$ coordinates, we use $\pd{r}{x}=\cos(\alpha)$, $\pd{r}{y}=\sin(\alpha)$,
% \begin{equation}
%     \pd{R}{x}=\cos(\alpha(\varphi,r)),\quad\pd{R}{x}=\sin(\alpha(\varphi,r))),
% \end{equation}
and 
% $ \pd{\varphi}{x} = -\frac{\sin(\alpha(\varphi,r)) + c\cos(\alpha(\varphi,r))} {\omega(1+r)}$, 
% $\pd{\varphi}{y} = \frac{\cos(\alpha(\varphi,r)) - c\sin(\alpha(\varphi,r))} {\omega(1+r)}$.
$ \pd{\varphi}{x} = -\frac{\sqrt{1+ \lambda^{2}}}{(r+1)\omega}\cos(\alpha - \psi)$, 
$ \pd{\varphi}{y} =  \frac{\sqrt{1+ \lambda^{2}}}{(r+1)\omega}\sin(\alpha + \psi)$, 
where $\psi = \tan^{-1}(\lambda)$ and $\lambda = c(r+1)H'(r)$.
% \begin{gather}
%   \label{eq:137}
%   \pd{\Phi}{x} =  \frac{cr\cos(\beta(\varphi, r))}{(r+1)(r+2)} - \frac{\sin(\beta(\varphi, r))}{\omega(r+1)}\\
% % c\left(\frac{2}{r+2} - \frac{1}{r+1}\right)\cos(\beta(\varphi, r)) - \frac{\sin(\beta(\varphi, r))}{\omega(r+1)}
% % \frac{2c\cos(\beta(\varphi, r))}{r+2} - \frac{1}{\omega(r+1)}(\sin(\beta(\varphi, r)) + c\omega\cos(\beta(\varphi, r)))
%   \pd{\Phi}{y} =  \frac{cr\sin(\beta(\varphi, r))}{(r+1)(r+2)} + \frac{\cos(\beta(\varphi, r))}{\omega(r+1)}
% \end{gather}
In phase-amplitude coordinates, \eqref{eq:14} becomes
\begin{align}
  \label{eq:1}
  \dot{\varphi}&= 1+\dif \en(\varphi, r) + \sigma \mathbf{h}(\varphi, r) \cdot \bm{\xi}(t)\\
  \dot{r}&=\gamma f(r) + \dif \tn(\varphi, r) + \sigma \mathbf{g}(\varphi, r) \cdot \bm{\xi}(t),\nonumber
\end{align}
where $h_{i}(\varphi, r) = G_{i}\pd{\varphi}{i}$, and $g_{i}(\varphi, r) = G_{i}\pd{r}{i}$, for $i=x,y$.
While the noise is additive in cartesian coordinates \eqref{eq:14}, it is multiplicative in $(\varphi, r)$ coordinates, and we use the Ito interpretation.
The terms that arise from the stochastic change of variables (i.e., the terms that come from the Ito calculus \cite{gardiner}) are $\en(\varphi, r) \equiv \mathbf{h}(\varphi, r)\cdot \pd{\mathbf{h}}{\varphi} + \mathbf{g}(\varphi, r)\cdot \pd{\mathbf{h}}{r}$ and $\tn(\varphi, r) \equiv \mathbf{h}(\varphi, r)\cdot \pd{\mathbf{g}}{\varphi} + \mathbf{g}(\varphi, r)\cdot\pd{\mathbf{g}}{r}$.
The term $\frac{\sigma^{2}}{2}\en(\varphi, r)$ can be directly interpreted as the average phase advance/delay due to noise during a single infinitesimal increment of the stochastic process and plays a significant role in the behavior of the process.
To see this, consider the arithmetic mean phase change due to perturbations $y_{c}\pm \hat{y}$ (for simplicity, assume the perturbations are to $y$ only, with $y_{c}$ the $y$ component of the limit cycle), given by
\begin{equation}
  \label{eq:19}
  \Delta \varphi(y_{c}, \hat{y}) \equiv \frac{1}{2}(\varphi(y_{c} + \hat{y}) + \varphi(y_{c} - \hat{y})) - \varphi(y_{c}).
\end{equation}
Expanding in a Taylor's series about $\hat{y}=0$, we find that $\Delta \varphi(y_{c}, \hat{y}) \sim \frac{\hat{y}^{2}}{2}\pdd{\varphi}{y} + O(\hat{y}^{4})$.
It follows that $\pdd{\varphi}{y} = \pd{h_{y}}{y} = (\pd{r}{y}\pd{h_{y}}{r} + \pd{\varphi}{y}\pd{h_{y}}{\varphi}) = (g_{y}\pd{h_{y}}{r} + h_{y}\pd{h_{y}}{\varphi})$, which is $\en(\varphi, r)$ for the case $G_{x}=0,\,G_{y}=1$.

The behavior of the oscillator can also be understood through terms that appear in the Fokker--Planck (FP) equation,
% \begin{equation}
%   \label{eq:9}
% \begin{split}
%   \pd{p}{t} &= -\pd{}{\varphi}(V_{\varphi}p) -\pd{}{r}(V_{r}p) + \dif \pd{}{\varphi}\left(D_{\varphi}\pd{p}{\varphi}\right) \\
% &\quad+ \dif \pd{}{r}\left(D_{r}\pd{p}{r}\right)   + \sigma^{2}\pd{}{\varphi}\left(g\cdot h \pd{p}{r} \right)
%  \end{split}
% \end{equation}
\begin{equation}
  \label{eq:10}
    \partial_{t}p = \sum_{i=\varphi,r}\partial_{i}\left[-v_{i}p + b_{i}\partial_{i}p\right] 
 + \frac{\sigma^{2}}{2}\sum_{\stackrel{i, j=\varphi,r}{i\neq j}}\partial_{i}(\mathbf{g}\cdot \mathbf{h}\, \partial_{j}p).
\end{equation}
We use conservation or Fickian form because it has the most natural connection to the stationary density.
The drifts are  $v_{\varphi}\equiv 1+\dif\left(\mathbf{g}\cdot \pd{\mathbf{h}}{r} - \mathbf{h}\cdot \pd{\mathbf{h}}{\varphi} \right)$,
$v_{r}\equiv \gamma f(r)+\dif\left(\mathbf{h}\cdot \pd{\mathbf{h}}{r} - \mathbf{g}\cdot \pd{\mathbf{h}}{\varphi} \right)$,
% \begin{gather}
%   \label{eq:3}
%   V_{\varphi}\equiv 1+\dif\left(\mathbf{g}\cdot \pd{\mathbf{h}}{r} - \mathbf{h}\cdot \pd{\mathbf{h}}{\varphi} \right)\\
%   V_{r}\equiv \gamma f(r)+\dif\left(\mathbf{h}\cdot \pd{\mathbf{h}}{r} - \mathbf{g}\cdot \pd{\mathbf{h}}{\varphi} \right),
% \end{gather}
and the diffusivities are $b_{\varphi} \equiv  \dif \abs{\mathbf{h}(\varphi, r)}^{2}$ and $b_{r} \equiv \dif \abs{\mathbf{g}(\varphi, r)}^{2}$.
It is convenient to refer to the drift and diffusivities on the limit cycle, so we define $V_{k}(\varphi) = v_{k}(\varphi, 0)$ and $D_{k}(\varphi) = b_{k}(\varphi, 0)$, where $k=\varphi,r$. 
% (Note that if $G_{x}=G_{y}$, the phase drift in \eqref{eq:1} is $1+\dif \en(\varphi, 0) = V_{\varphi}$.)
% The Fokker--Planck equation corresponding to \eqref{eq:1} is
% \begin{equation}
%   \label{eq:2}
%   \pd{p}{t} = -\pd{}{\varphi}\left(V_{\varphi}( \right) - \pd{}{r}\left( \right)
% \end{equation}

If the system is rotationally symmetric ($G_{x}=G_{y})$, the drifts and diffusivities are independent of $\varphi$.
In this case, the (FP) equation is
\begin{equation}
  \label{eq:8}
  \pd{p}{t} = -\pd{}{r}(v_{r}(r) p)  + \dif \pdd{p}{r}, \quad   v_{r} = \gamma f(r) + \dif \frac{1}{r+1}.
\end{equation}
The stationary density is 
\begin{equation}
  \label{eq:7}
  p_{ss}(\varphi, r) = N(r+1)\explr{-\frac{\gamma}{2\sigma^{2} } r^{2}(r+2)^{2}},
\end{equation}
where $N$ is a normalization constant.
% Using Laplace's method we have
% \begin{equation}
%   \label{eq:11}
%   N \sim 
% \end{equation}
The marginal stationary density for the phase is uniform, that is $u_{ss}(\varphi) \equiv \int_{-1}^{\infty}p_{ss}(\varphi, r)dr = \frac{1}{2\pi}$.
% \begin{equation}
%   \label{eq:16}
%   u_{ss}(\varphi) \equiv \int_{-1}^{\infty}p_{ss}(\varphi, r)dr = \frac{1}{2\pi}.
% \end{equation}
Hence, the trajectories are approximately gaussian distributed with respect to $r$ and uniformly distributed in phase.

\begin{figure}[htb]
  \centering
  \includegraphics[width=8cm]{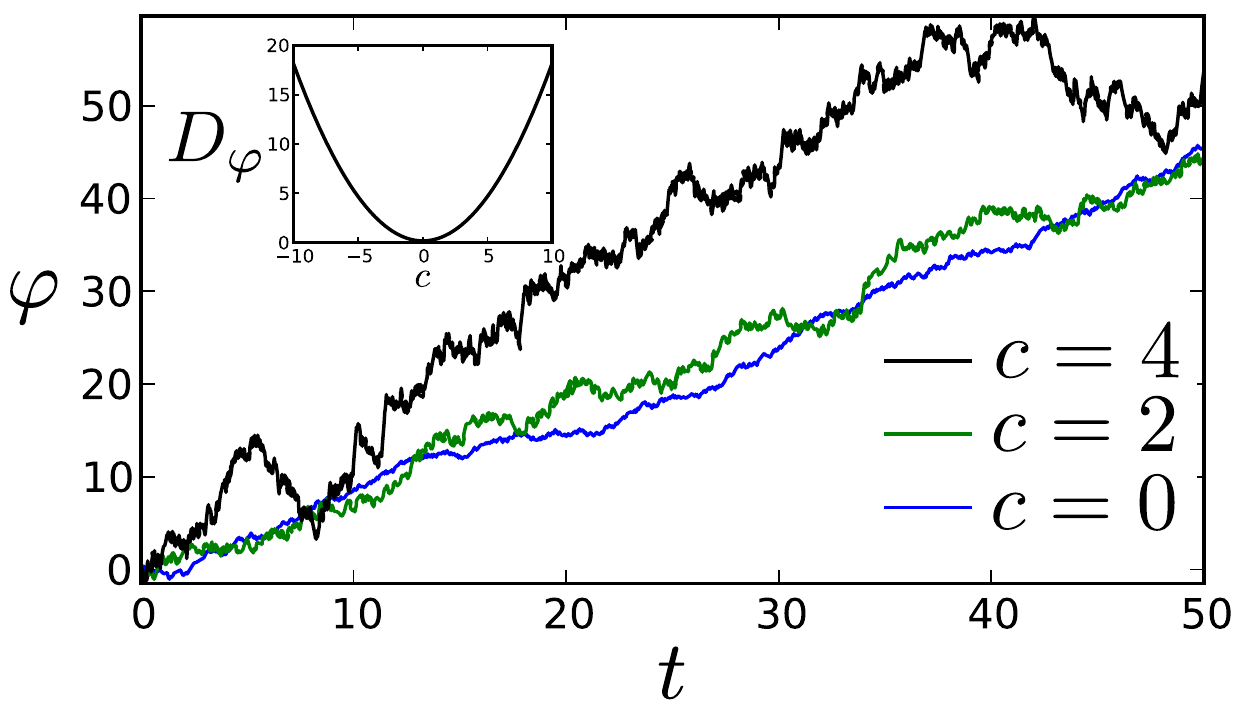}
  \caption{The counter-rotating case (see Fig.~\ref{fig:cases}a,b), with $G_{x}=G_{y}$. Parameter values are $\gamma=15$, $\omega=1$, and $\sigma=0.6$.} 
  \label{fig:symV1}
\end{figure}

For the counter-rotating case (Fig.~\ref{fig:cases}a,b), $V_{\varphi}=1$ is independent of $c$.
Hence, the average period is approximately $2\pi/\omega$.  
On the other hand, $D_{\varphi} =  \frac{\sigma^{2}(c^{2} + 1)}{2\omega^{2}}$, and it follows that as the magnitude of $c$ is increased, the fluctuations in phase are amplified like $c^{2}$ (see Fig.~\ref{fig:symV1}).
% The reason for this is as follows. 
From Fig.~\ref{fig:cases}a,b, we see that when a deterministic oscillator is perturbed across $\rho_{*}$ (dashed line) its phase decreases.  
Otherwise, the phase increases as it rotates along with the limit cycle.
% From Fig.~\ref{fig:cases}(a,b), we see that when a deterministic oscillator is perturbed toward $\rho_{*}$ (dashed line) its phase is delayed.  
% On the other hand, perturbations away from $\rho_{*}$ cause a phase advance.
The effect of noise pushing the oscillator in both directions is a random circular-like motion that effectively amplifies the fluctuations in phase without affecting the average frequency of oscillations.

\begin{figure}[htb]
  \centering
  \includegraphics[width=8cm]{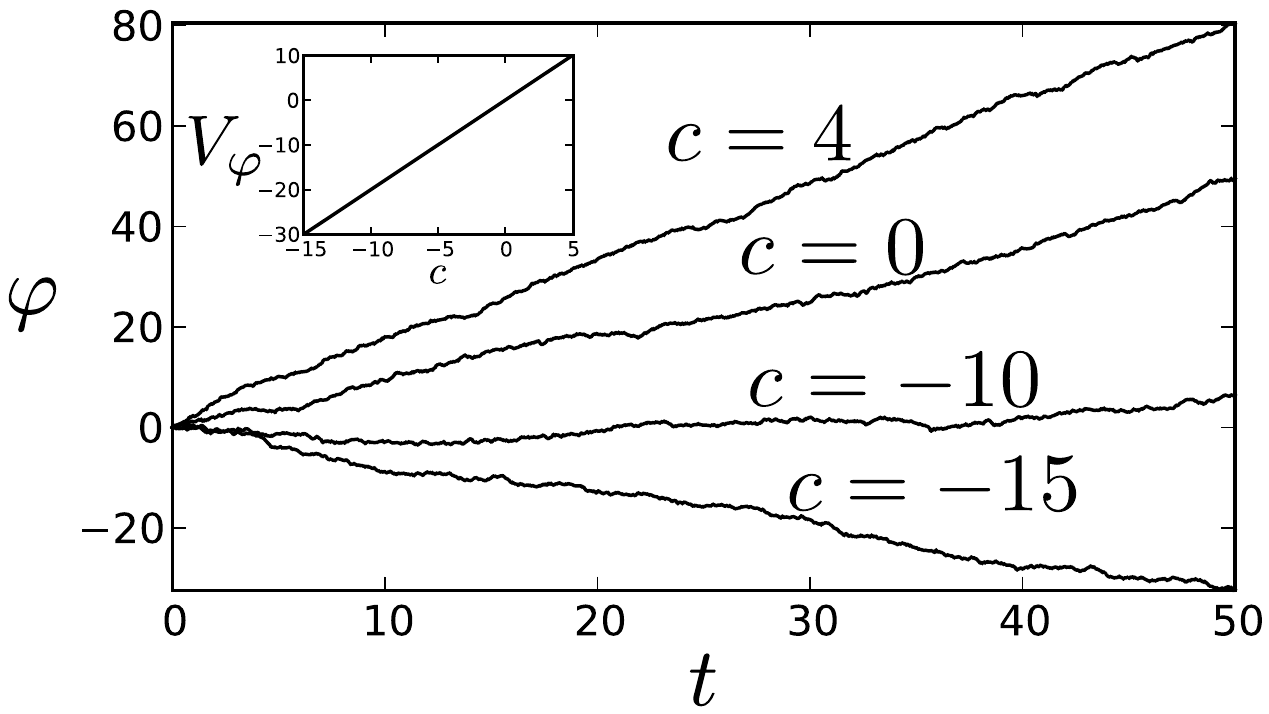}
  \caption{The uniformly-rotating and anti-rotating cases (see Fig.~\ref{fig:cases}c,d), with $G_{x}=G_{y}$. Parameter values are $\gamma=15$, $\omega=1$, and $\sigma=0.63$.}
  \label{fig:symV2}
\end{figure}

The situation is reversed for the anti-rotating case (Fig.~\ref{fig:cases}d); $D_{\varphi} = \frac{1 }{\omega}$ is independent of $c$, while $V_{\varphi} = 1 + \frac{\sigma^{2}c}{2}$ is a linear function of $c$.  
Hence, for $c > 0$ the average frequency increases and for $c<0$ it slows down, while the noise in phase is unaffected (see Fig.~\ref{fig:symV2}). 
It follows that when $c = c_{0} = -8/\sigma^{2}$, the oscillator stops oscillating and behaves like brownian motion on a periodic domain.
Finally, for $c<c_{0}$ the stochastic oscillator rotates in the opposite direction of the deterministic limit cycle.

Without rotational symmetry, \eqref{eq:7} is no longer valid and the marginal density $u_{ss}$ is no longer uniform, but the joint stationary density is still approximately Gaussian in $r$.
To see what happens when the rotational symmetry is broken (i.e.,  $G_{x}\neq G_{y}$), we use an averaging approximation, which projects the dynamics onto the limit cycle by averaging out $r$ to get single SDE for the phase \cite{yoshimura2008,teramae2009}.
If the limit cycle is strongly attracting then \eqref{eq:1} is, to leading order in $\gamma^{-1}$, approximated by the Ito-type SDE,
\begin{equation}
  \label{eq:4}
  \dot{\varphi}\sim 1+\dif \en_{0}(\varphi) + \sigma h_{0}(\varphi) \xi(t), \quad \gamma \gg 1,
\end{equation}
where $\en_{0}(\varphi) \equiv \en(\varphi, 0)$ and $h_{0}(\varphi) \equiv h_{x}(\varphi, 0) + h_{y}(\varphi, 0)$.
The stationary density is approximately
\begin{equation}
  \label{eq:17}
  p_{ss}(\varphi, r) \sim u_{ss}(\varphi)N\explr{-\frac{\gamma r^{2}}{2D_{r}(\varphi)}},%;% normalization constant
% u_{ss}(\varphi) = \explr{\int_{0}^{\varphi}\frac{1+\dif n_{0}(s)}{\dif\Vert \mathbf{h}_{0}(s)\Vert^{2}}ds}
\end{equation}
where $u_{ss}(\varphi)$ is the marginal density, satisfying $\partial_{\varphi}\left(D_{\varphi}(\varphi)\partial_{\varphi}u_{ss} - V_{\varphi}(\varphi) u_{ss}\right) = 0$.

\begin{figure}[htb]
  \centering
  \includegraphics[width=8cm]{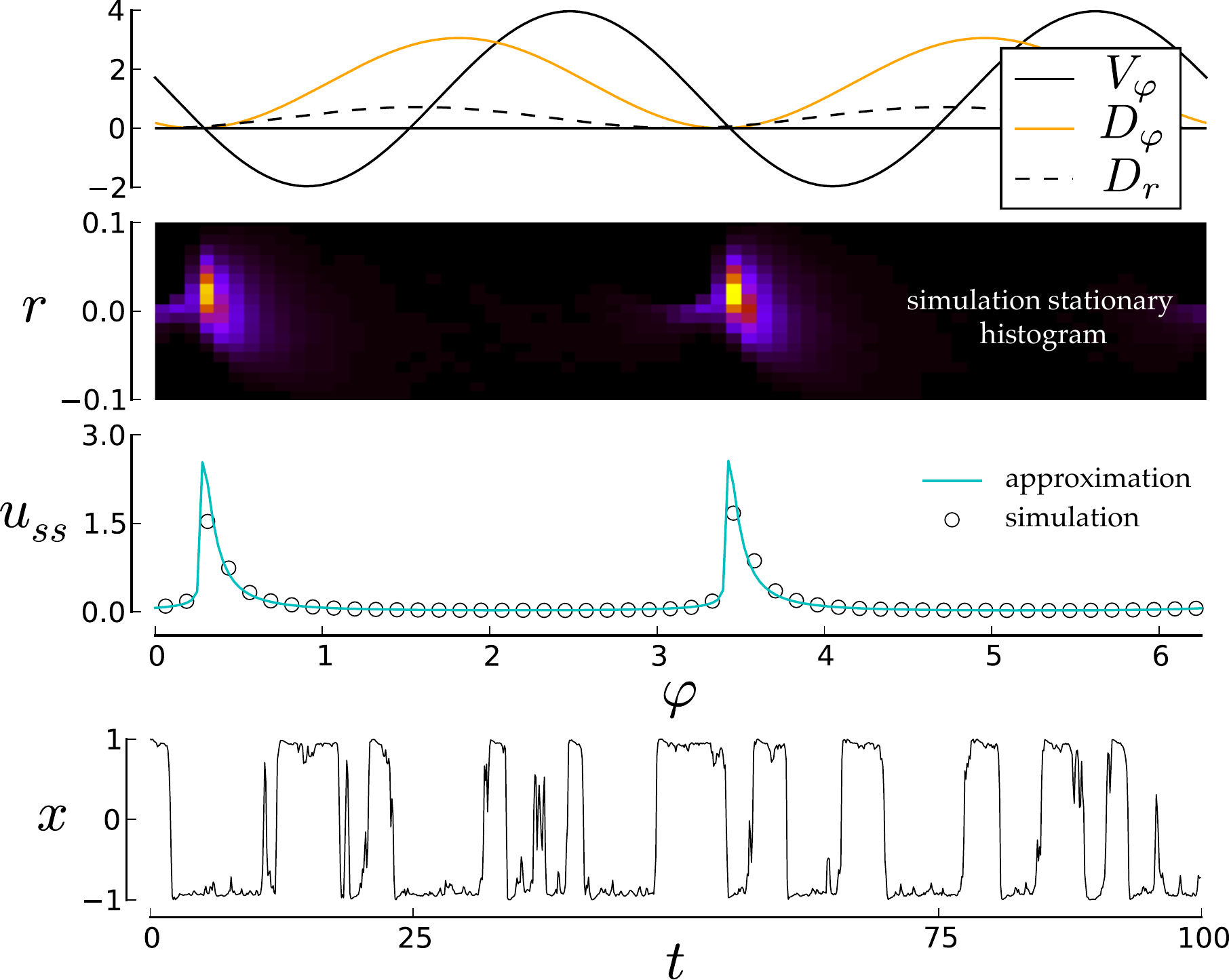}
  \caption{Broken symmetry for the counter-rotating case (see Fig.~\ref{fig:cases}a).  $c=4$, $\gamma = 15$, $\sigma = 0.6$, $G_{x} = 0$, and $G_{y} = 1$.  $x(t)$ is a simulation trajectory showing bistable-like behavior.  The stationary density is sharply peaked at the zeros of the drift $V_{\varphi}$. Note that the zeros of $D_{\varphi}$ are not exactly at the zeros of $V_{\varphi}$.}
  \label{fig:nosymV1}
\end{figure}

% If $G_{x}\neq G_{y}$ the system is no longer has statistical radial symmetry.
For simplicity, we take $G_{x}=0$ and $G_{y}=1$ so that noise is only in the $y$ direction.
In this case, $V_{\varphi}$ and $D_{\varphi}$ are functions of $\varphi$.
Bistability occurs for moderate values of $\sigma$; there are two {\em turning} points where $V_{\varphi}$ changes sign from positive to negative (see Fig.~\ref{fig:nosymV1}) as $\varphi$ is increased. 
The turning points are local maxima of the stationary density, where the oscillator spends most of its time, struggling to move past a dynamical barrier imposed by the noise. 
Note that without rotational symmetry, the Ito drift from the SDE \eqref{eq:4} is not the same as the Fickian drift, $V_{\varphi}$, from the FP equation \eqref{eq:10} (with rotational symmetry the drift terms are the same because $\partial_{\varphi}\mathbf{h} = 0$).
This is because, in general, the average increment of the process during an infinitesimal time $dt$ is not directly related to the stationary density.
That is, the former determines the average direction of the next infinitesimal jump at a given point, while the latter is the relative fraction of time the process spends in a neighborhood of that point as $t\to\infty$.
Bistability is a phenomena that is defined in terms of the stationary density (i.e., that the process spends the vast majority of time near two well separated points, which results in a stationary density with two sharp peaks, see Fig.~\ref{fig:nosymV1}).
This means that in general, one cannot infer bistability by thinking about the average phase change due to a fluctuation.
% The connection must be translated from the Ito drift to the Fickian drift through the FP equation.

The standard example of a bistable system perturbed by noise is diffusion in a double well potential, known as Kramers' problem.
In Kramers problem, the frequency of transitions between the two minima of each well is exponentially decreasing as the noise magnitude goes to zero so that without noise, such transitions do not occur.
However, the stochastic oscillator becomes {\em more} bistable (the transitions between the two metastable phases become {\em less} frequent) as the noise magnitude is {\em increased}.

\begin{figure}[tb]
  \centering
  \includegraphics[width=8cm, height = 6cm]{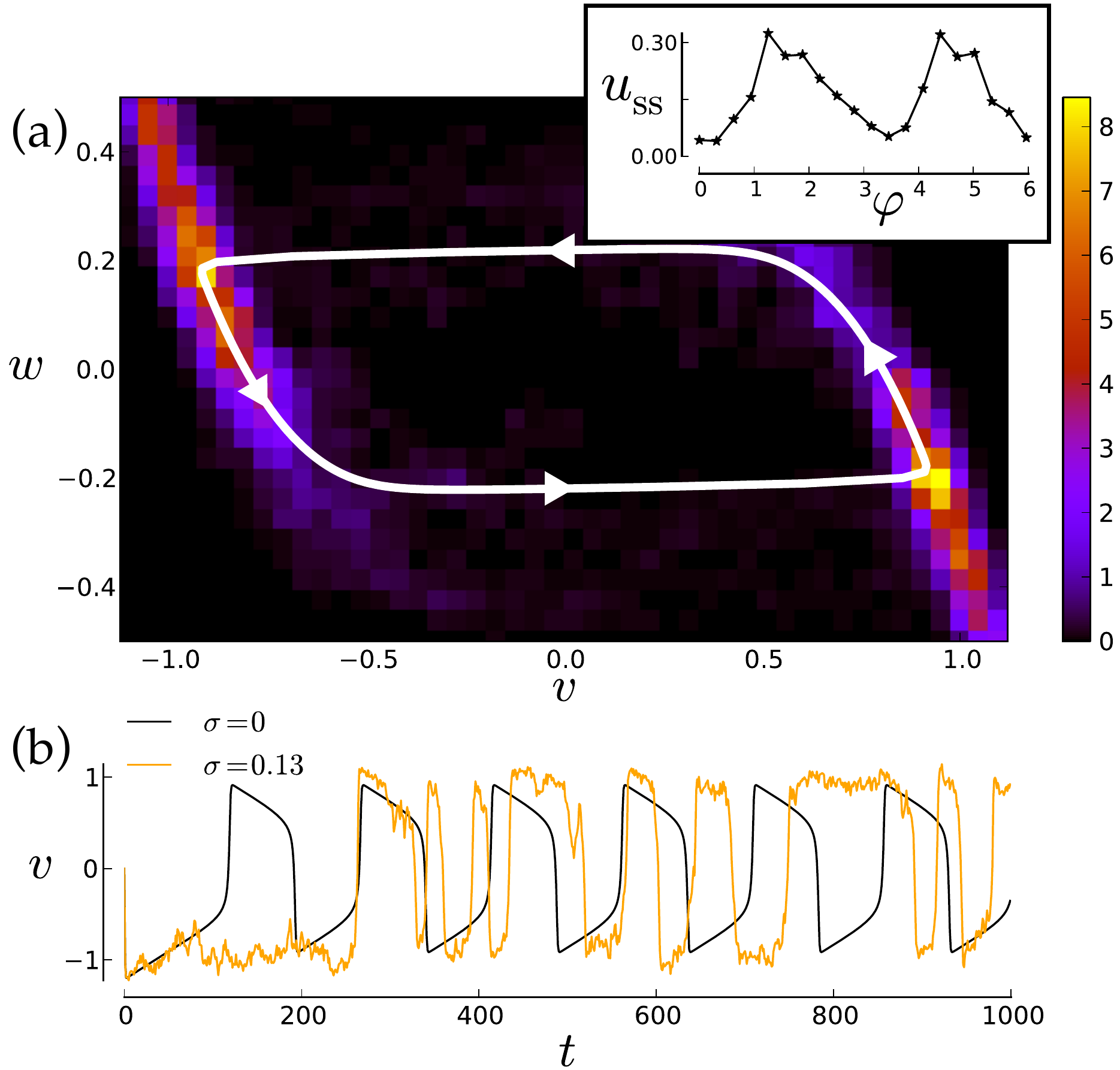}
  \caption{Bistability in the FHN model. (a) Phase plane with stationary histogram, for $\sigma = 0.13$. (The inset shows the marginal histogram of asymptotic phase.) Thick white line shows the deterministic limit cycle.
% and thin white lines show isochrones.
 (b) Sample stochastic trajectories. Parameters are $\tau = 100$, $a = 0.8$, and $m = 1.2$.}
  \label{fig:fhn}
\end{figure}

To verify that bistability can occur in more complex oscillators, we show simulation results for the FHN oscillator \cite{fitzhugh1961,nagumo1962},
\begin{equation*}
  \label{eq:15}
   \dot{x} = -x(x^{2} - a^{2}) - y + \sigma \xi_{x}(t),\quad
 \dot{y} = \frac{x - m y}{\tau} + \sigma \xi_{y}(t).
\end{equation*}
% \begin{align}
%   \label{eq:13}
%  \dot{x} &= -x(x^{2} - a^{2}) - y + \sigma G_{x}\xi_{x}(t), \\\nonumber
%  \dot{y} &= (x - m y)/\tau + \sigma G_{y}\xi_{y}(t).
% \end{align}
As shown in Fig.~\ref{fig:fhn}, sample trajectories are qualitatively bistable-like, and the stationary histogram is sharply peaked near two locations on the limit cycle.  
% Near the left most peak of the histogram, positive fluctuations in $w$ cause a large phase delay [see isochrones (thin white lines) in Fig.~\ref{fig:fhn}(a)] while negative fluctuations cause a smaller (in magnitude) phase advance.
The phase reduction of the FHN model is more complicated than for the radially-symmetric oscillators, and this analysis will be presented elsewhere.  
However, in the inset of Fig.~\ref{fig:fhn}(a), we show that the histogram of the numerically computed asymptotic phase displays two distinct peaks.

We have shown that a simple dynamical system with a limit cycle can have qualitatively different behavior when moderate noise is present.
In particular, the system can appear bistable, rotate in the opposite direction of the deterministic limit cycle, or cease oscillating altogether.
% If the process has statistical rotational symmetry, a mechanistic explanation of the noise induced dynamics can be inferred from the deterministic vector field.
We have also shown that, in general, it is difficult to infer bistability from the underlying deterministic system.
However, if the limit cycle is strongly attracting, then phase reduction can be used to predict the emergence of bistable behavior.
Nevertheless, we are able to demonstrate that bistable behavior is also observed in the more complex FHN oscillator.

 Noise induced effects may lead to interesting behavior for coupled oscillators, in particular, oppositely rotating coupled oscillators (see, for example, \cite{bhowmicketal2011,prasad2010}).
%Recently, it has been shown that diffusively coupling oppositely-rotating oscillators can lead to a type of mixed synchronization where complete synchronization and antisynchronization coexist in different state variables \cite{bhowmicketal2011,prasad2010}.
% To observe this effect, the underlying dynamical system of an oscillator had to be altered in order to obtain opposite rotations prior to coupling.
% Our results suggest that coupled oppositely-rotating oscillators can also be achieved by adding different levels of noise to identical oscillators.
% It would be interesting to explore whether the same synchronization phenomena occur in this case.
% This could lead to interesting synchronization phenomena with applications to counter-rotating vortices in fluid mechanics \cite{bhowmicketal2011,meunierandleweke2005}.
Our analysis could also be extended to include colored noise as discussed in \cite{teramae2009}.
The correlation time has been shown to affect the noise induced drift term, which would then influence the results discussed here.

%Merlin.mbs v4.21 2009-07-09.
%

% \bibliography{Refs}
\end{document}